# Interface Control of Second-Harmonic Generation in Layered $CuInP_2S_6$ via $PbZr_{0.2}Ti_{0.8}O_3$ Domain Wall

Dawei Li,[1,2]* Juhe Liu,[1,2] Yifei Hao,[3] Qiuchen Wu,[3] Xi Huang,[4] Jesse Stebbing,[3] Xinyi Hou,[1] Gaolei Zhao,[1] Kun Wang,[3] Yongfeng Lu,[4] and Xia Hong[3]*

[1] School of Optoelectronic Engineering and Instrumentation Science, Dalian University of Technology, Dalian, Liaoning 116024, China.

[2] Dalian University of Technology and Belarusian State University Joint Institute, Dalian University of Technology, Dalian, Liaoning 116024, China

[3] Department of Physics and Astronomy, University of Nebraska-Lincoln, Lincoln, NE 68588-0511, United States

[4] Department of Electrical and Computer Engineering, University of Nebraska-Lincoln, Lincoln, NE, 68588-0511, United States

* Correspondence to dwli@dlut.edu.cn, xia.hong@unl.edu

## Abstract

The control of nonlinear optical responses in van der Waals ferroelectrics at the nanoscale offers opportunities for developing two-dimensional integrated nanophotonics with programmable functionalities. In this work, we report the first realization of nanoscale nonvolatile control of the amplitude and polarization of second-harmonic generation (SHG) response in layered $CuInP_2S_6$ (CIPS) via the domain structure in an interfacial $PbZr_{0.2}Ti_{0.8}O_3$ (PZT) thin film. The SHG intensity at CIPS/PZT heterointerface is continuously modulated, either substantially enhanced or almost entirely quenched, by the circular domain walls in underlying ferroelectric oxide. The interfacial coupling of CIPS polar axis with in-plane polarization of PZT domain walls tailors the SHG light polarization into two-fold ($C_{2V}$) symmetry, which is well modeled by nonlinear electromagnetic theory. Overall, this study enables nonvolatile programmable design of nonlinear light intensity and polarization in van der Waals ferroelectrics at the nanoscale.

Ferroelectrics, with the nonvolatile, electrically switchable polarization, have been widely exploited to develop energy efficient memory, neuromorphic computing, etc.[1,2] In recent years, two-dimensional (2D) van der Waals (vdW) ferroelectrics have attracted intensive research interest as they can retain robust ferroelectricity down to the monolayer limit.[3,4] The highly tunable, stackable nature and strong nonlinear optical response of 2D ferroelectrics make them a versatile platform for exploring light-matter interaction.[5] Various novel functionalities have been discovered at 2D ferroelectrics and 2D ferroelectric-based heterostructures, including ferroelectric tunneling junction states,[6,7] memristive effects,[8,9] ferroelectric field effects,[10,11] negative capacitance effects,[12] and tunable photoresponses.[13,14] Previous studies have focused on exploiting the polarization of 2D ferroelectrics to tune the electronic and optoelectronic properties of interfacial 2D semiconductors, while the potential for designing their nonlinear optical response is yet to be explored.

As 2D ferroelectrics are polar materials, they are ideal material platforms for exploring various nonlinear optical phenomena. In previous studies, second-harmonic generation (SHG),[15-19] sum-/difference-frequency generation,[20,21] third-harmonic generation,[22] and four-wave mixing[17,20] have been observed in a wide range of polar vdW materials, including $MoS_2$, $MoTe_2$, and *h*-BN. Recent studies have revealed strong nonlinear optical response in 2D ferroelectrics. For example, $NbOX_2$ (X = Cl, Br, I), which possesses in-plane polarization, exhibits strong and anisotropic optical SHG responses.[23-27] α-$In_2Se_3$, which exhibits intercorrelated in-plane and out-of-plane polarization,[28] shows SHG responses with high stability.[29,30] $CuInP_2S_6$ (CIPS) is known for its relatively high Curie temperature and large out-of-plane polarization.[31,32] Strong SHG response has been observed in CIPS due to its noncentrosymmetric crystal structure [*Cc*, Fig. 1(a)].[33-35] The nanoscale, nonvolatile control of SHG response in CIPS and other 2D ferroelectrics is critical for their implementation in nanophotonics with programmable functionalities, which has not been realized to date.[34,36-38]

In this work, we report the giant and tunable SHG response in layered CIPS and demonstrate for the first-time nanoscale control of the SHG signal in CIPS of various thicknesses through integration with ferroelectric $PbZr_{0.2}Ti_{0.8}O_3$ (PZT) thin films. By interfacing few-layer to thick CIPS flakes with prepatterned circular domains, we achieve spatial control of the magnitude and polarization of the SHG signal at domain walls (DWs). The SHG intensity at the heterointerface can be continuously modulated by the in-plane polarization of the underlying circular DW. In addition, the SHG light polarization exhibits two-fold symmetry ($C_{2v}$) pattern at the PZT DW, which has been modeled by nonlinear electromagnetic theory. The CIPS/PZT heterostructure thus provides a powerful material platform for achieving nanoscale programing of nonlinear light intensity and polarization, which is promising for developing smart optical modulation and optical information processing.

Ferroelectric CIPS features a vdW layered structure of non-centrosymmetric monoclinic symmetry, space group $C_C$, [Fig. 1(a)][33] where the S atoms constitute the octahedral network within each layer, caging the cations ($Cu^+$, $In^+$) and P-P units. The CIPS unit cell has the lattice parameters of $a$ = 6.096 Å, $b$ = 10.565 Å, $c$ = 13.623 Å, and $\beta$ = 107.10°, and contains two adjacent layers.[39] CIPS flakes are mechanically exfoliated from the bulk crystals (Supplemental Materials) and transferred onto the target substrates (*e.g.*, $SiO_2$/Si, gold, and PZT).[40] Thinner flakes are identified optically and their thicknesses are determined by atomic force microscopy (AFM) measurements. Figure 1(b) shows the Raman spectrum of a CIPS flake (~106 nm) on $SiO_2$/Si substrate. Several Raman active modes are observed, including 100-137 $cm^{-1}$ for the anion, 159 $cm^{-1}$ for S-P-P bond [δ(S-P-P)], 233-297 $cm^{-1}$ for S-P-S bond [δ(S-P-S)], 314 $cm^{-1}$ for the cation, and 377 $cm^{-1}$ for P-P stretching [$\nu$(P-P)], which are in good agreement with previous reports of ferroelectric CIPS.[41,42] Figures 1(c)-1(d) show the piezoresponse force microscopy (PFM) switching hysteresis of phase and amplitude responses taken on a CIPS flake (~16 nm) transferred onto the gold substrate, where the coercive voltages for polarization down ($V_{tip} > 0$ V) and up ($V_{tip} < 0$ V) are 4.5 V and -5.2 V, respectively.

Next, we investigate the SHG responses in CIPS using a home-made nonlinear optical microscope (Supplemental Materials).[40] The femtosecond laser (wavelength: 800 nm) as an incident light propagates along $z$ direction, perpendicular to the sample surface, and the SHG signal (wavelength:

400 nm) is collected in reflection mode. Figure 1(e) shows the SHG mapping of an exfoliated CIPS flake with a thickness of about 50 nm, where the incident light polarization is along *x* direction and no analyzer is applied. The CIPS flake exhibits a strong SHG signal with spontaneous domain structures. Previous studies have shown that CIPS exhibits a thickness-dependent transition from the monoclinic (*Cc*) to trigonal (*P*31*c*) phase, with the former possessing in-plane polarization.[43] To probe the crystal orientation and structure symmetry of CIPS, we have performed polarization dependent SHG measurements. Figure 1(f) shows the parallel polarized SHG mapping of the same CIPS sample shown in Fig. 1(e) (red-circled region) taken at different rotation angles $\varphi$ ($0^{\circ} \leq \varphi \leq 60^{\circ}$). The SHG signal reaches maximum intensity at $\varphi \approx 0^{\circ}$ and $60^{\circ}$, and minimum intensity at $\varphi \approx 30^{\circ}$. The parallel polarized SHG intensity versus rotation angle $\varphi$ exhibits a six-fold petal pattern [Fig. 1(g)],[33] which can be modeled using the nonlinear electromagnetic theory. The intensity of polarized SHG signal is given by:[44]

$$I_{\mathrm{SHG}} = \left|\mathbf{P}_{\mathrm{CIPS}}^{2\omega}\right|^2 \propto \left|\hat{e}_{2\omega} \cdot d_{\mathrm{CIPS}}^{(2)} \cdot \hat{e}_{\omega}^2\right|^2, \tag{1}$$

where $\mathbf{P}_{\mathrm{CIPS}}^{2\omega}$ is the polarization vector of the SHG light, $\hat{e}_{\omega} = \hat{e}_{2\omega} = [\cos(\varphi), -\sin(\varphi), 0]$, $d_{\mathrm{CIPS}}^{(2)}$ is the second-order *d*-tensor for CIPS. For monoclinic CIPS with *m*-point group symmetry, there are ten nonvanishing elements ($d_{11}$, $d_{12}$, $d_{13}$, $d_{15}$, $d_{24}$, $d_{26}$, $d_{31}$, $d_{32}$, $d_{33}$, and $d_{35}$) and six independent elements in $d_{\mathrm{CIPS}}^{(2)}$.[37,45] Thus, $d_{\mathrm{CIPS}}^{(2)}$ can be expressed as:

$$d_{\mathrm{CIPS}}^{(2)} = \begin{bmatrix} d_{11} & d_{12} & d_{13} & 0 & d_{15} & 0 \\ 0 & 0 & 0 & d_{24} & 0 & d_{26} \\ d_{31} & d_{32} & d_{33} & 0 & d_{35} & 0 \end{bmatrix}, \tag{2}$$

where $d_{26} = d_{12}$, $d_{31} = d_{15}$, $d_{32} = d_{24}$, and $d_{35} = d_{13}$. In our case, $d_{11} \approx d_{12}$ and $d_{15} \approx d_{24}$.[37] By substituting Equation (2) into Equation (1), we obtain the parallel polarized SHG intensity ($I_{\mathrm{SHG},\parallel}$) as:

$$I_{\mathrm{SHG},\parallel} = I_0 \cos^2(3\varphi) \propto d_{11} \cos^2(3\varphi), \tag{3}$$

where $I_0$ is the maximum intensity of the SHG response. The modeled SHG intensity result using Equation (3) [blue curve in Fig. 1(g)] shows excellent agreement with the experimental data [black dots in Fig. 1(g)]. Based on its lattice symmetry [Fig. 1(a)], like monolayer (1L) 1H-$MoS_2$,[44] the SHG electric field ($\mathbf{P}^{2\omega}$) versus CIPS crystal orientation exhibits three-fold rotational symmetry. We thus can establish the relationship between $I_{\mathrm{SHG},\parallel}$ and CIPS crystal orientation: the maximum (minimum) SHG intensity is parallel (perpendicular) to one of polar axes of CIPS [pink arrows in Fig. 1(g)].

Our previous studies have revealed the complex interface interaction between CIPS and ferroelectric PZT mediated by strain and light adsorption.[32,46] While the polar alignment between PZT and vdW layer can modify the interfacial symmetry and the associated second-order dielectric tensor,[44,47] its impact on the SHG response at the CIPS/PZT heterointerface remain unexplored. Next, we prepare the CIPS/PZT heterostructures and investigate their SHG responses [Fig. 2(a)]. We deposit epitaxial 50 nm (001) PZT thin film on 10 nm $La_{0.67}Sr_{0.33}MnO_3$ (LSMO) buffered $SrTiO_3$ (STO) substrate (step 1). The PZT films are smooth with a typical root-mean-square (RMS) surface roughness of ~2 Å (Supplemental Materials).[40] Circular domain patterns are then written on the PZT film via conductive AFM tip (step 2). The as-deposited PZT thin film is *c*-axis orientated with out-of-plane polarization uniformly polarized in the $P_{\mathrm{up}}$ state.[47] Figures S3(b) and S3(c) show the PFM amplitude and phase images of a circular domain written on PZT film. The region outside of the circle is uniformly polarized in the $P_{\mathrm{up}}$ state (or along [001] direction), while the region inside circle is polarized in the $P_{\mathrm{dn}}$ state (or along $[00\bar{1}]$ direction), as confirmed by $180^{\circ}$ phase change (Supplemental Materials).[40] The domain wall (DW) position is clearly visible in the PFM amplitude image (Supplemental Materials).[40] It has been reported that CIPS exhibits excellent SHG performance from a few-layer to a bulk-like dimension.[34] In step 3, we select CIPS flakes of various thicknesses (9-50 nm) and transfer them on top of the patterned domain region. For thick CIPS flakes, we determine the polar axis via polarized SHG [Fig. 1(g)] and align it along the edge of the PZT film (<100> orientation) during transfer. For thinner CIPS flakes with diminishing SHG signal, the polar axis cannot be determined prior to transfer.

Figure 2(b) shows the AFM topography image of a circular domain on PZT with a ~9 nm CIPS flake transferred on top, forming thin CIPS/PZT heterostructure. The PFM amplitude [Fig. 2(c)] and phase images [Fig. 2(d)] show a circular domain structure on PZT, fully conforming to that on bare PZT. The PFM amplitude signal of CIPS/PZT stack is weaker than that of bare PZT [Fig. 2(c)], with the suppressed signal at the DW clearly visible. Our previous study has shown that the strain coupling between PZT and interfacial CIPS tilts the quadruple well, rendering full polar alignment in ultrathin CIPS flakes (below 25 nm) in direction opposite to PZT polarization.[46] The net piezoresponse yields a domain structure consistent with that of underlying PZT, which is corroborated by the 180$^o$ change in the PFM phase [Fig. 2(d)]. Similar polar alignment has been observed in a ~16 nm CIPS flake transferred on PZT domain [Figures 2(e)-2(g)]. For thicker CIPS (above 50 nm), the interfacial effect decays and spontaneous domain structures start to form. Figure 2(h) shows the AFM topography image of the same region as in Supplemental Materials after a ~50 nm CIPS flake transferred on top. PFM amplitude [Fig. 2(i)] and phase [Fig. 2(j)] images reveal emergence of nanoscale $P_{\mathrm{up}}$ or $P_{\mathrm{dn}}$ domains with length scale of 100 nm on top of the circular domain structure, with much weaker amplitude signal [Fig. 2(i) and Supplemental Materials].[40] These features are consistent with those observed on CIPS flakes on non-polar substrates,[46] suggesting the decayed interfacial synergy with underlying PZT.

Next, we carry out SHG measurements [Fig. 3(a) and Supplemental Materials] on the CIPS/PZT samples shown in Fig. 2. Figure 3(b) shows the SHG image of a bare PZT, where SHG response has only been detected on the circular DW, suggesting the existence of an in-plane polarization ($\vec{P}_{\mathrm{DW}}$) on a Née l-type DW.[48] The width of the detected SHG signal at the DW is about 500 nm, approaching the resolution of the SHG microscopy system. Our previous study reveals that the orientation of $\vec{P}_{\mathrm{DW}}$ [dashed yellow arrows in Fig. 3(b)] is normal to the DW direction (or along the radial direction).[44]

Figure 3(c) shows the SHG image of the same PZT domain in Fig. 2(h) after a thick CIPS flake (~50 nm) transferred on top, taken with incident light polarization (red solid arrow) along *y*-axis (*a*- axis of CIPS) and no analyzer applied. In the regions on $P_{up}$ and $P_{dn}$ domains, we observe the strong and comparable SHG signal from CIPS. At the DW region, interestingly, the SHG intensity of 50 nm CIPS is spatially modulated by the $\vec{P}_{DW}$ orientation $\theta$ with respect to polar *b*-axis. Unlike the isotropic angular dependence of SHG signal at DW for bare PZT [Fig. 3(b)], we observe highly enhanced SHG response at DW with $\theta \approx 0^o$ and significantly suppressed SHG response at DW with $\theta \approx 180^o$. With increasing $\theta$, the amplitude of the SHG response is modulated continuously, first decreasing for $0^o < \theta < 180^o$, reaching the minima at $\theta = 180^o$, and then increasing for $180^o < \theta < 360^o$. Figure 3(d) shows the SHG image of the 16 nm CIPS flake shown in Fig. 2(e), where the polar *b*-axis is along *y*-direction. As expected, significantly weaker SHG responses are observed on $P_{up}$ and $P_{dn}$ domains due to vanishing in-plane polarization.[46] Despite the much weaker SHG signal at the DW due to the thinner CIPS thickness, the spatial modulation of SHG response is clearly visible and follows the same angular dependence as that of the 50 nm CIPS. Figure 3(e) shows the SHG image of the 9 nm CIPS flake shown in Fig. 2(b), where the polar *b*-axis is along *x*-direction. While the SHG response on the $P_{up}$ and $P_{dn}$ domains are fully quenched, the SHG intensity at the DW regions exhibits similar spatial modulation. For comparison, Fig. 3(f) shows the polar plots of normalized SHG intensity for all samples. Here, we define the angle of $\theta'$ as the orientation of SHG signal at DW, where $\theta' = 0^o$ when SHG signal reaches maximum. It clearly illustrates the highly consistent angular-dependence of SHG tuning effect at CIPS/PZT heterointerfaces in contrast to the isotropic response on bare PZT. In addition, for all CIPS/PZT samples with various CIPS thicknesses [Fig. 3(g)], no significant difference in the SHG signal is observed between the $P_{up}$ and $P_{dn}$ domains, confirming that the SHG response in CIPS is unaffected by the PZT domains and is solely modulated by the PZT DWs.

The SHG evolution occurred at CIPS/DW heterostructure regions could be explained by interfacial polar coupling [Fig. 3(f) inset]. The in-plane polar structure of CIPS is similar to that of 1L 1H-$MoS_2$, which has three polar axes [pink arrows, $\theta = 60^o$, $180^o$, and $300^o$ in Fig. 3(c) inset]. For DW ($\theta \approx 0^o$), $\vec{P}_{\mathrm{DW}}$ is perpendicular to the incident light polarization as well as parallel to one of the polar axes of CIPS [$\vec{P}_{\mathrm{CIPS}}(\theta \approx 180^{\mathrm{o}})$]. In this case, the interfacial SHG dipole fields between CIPS and DW are coherently coupled, which leads to the maximum SHG intensity. For DW ($\theta \approx 180^{\mathrm{o}}$), whose $\vec{P}_{\mathrm{DW}}$ is anti-parallel to $\vec{P}_{\mathrm{CIPS}}(\theta \approx 180^{\mathrm{o}})$, the SHG dipole fields between CIPS and DW cancel each other, leading to the minimum SHG signal. The relative orientation between the polar axes of CIPS and PZT DW continuously tune the net SHG intensity, resulting in a highly consistent spatial modulation of SHG response shown in Fig. 3(f).

To gain a quantitative understanding of the tunable SHG response in CIPS/PZT heterostructures, we investigate the effect of incident light polarization on the SHG response in PZT and CIPS/PZT heterostructures. Supplemental Figure S5 shows the SHG images of bare PZT film in Fig. 3(b) taken with different incident light polarization angles $\varphi$ (relative to *y*-axis) and no analyzer applied. As expected, the SHG response is uniform along the circular DW, independent of incident light polarization. Figures 4(a)-4(d) shows the SHG images of the 50 nm CIPS/PZT heterostructure [Fig. 3(c)] taken at different incident light polarization angles $\varphi$ with no analyzer applied, where the incident light polarization has a clear impact on the spatial SHG distribution. The spatial SHG intensity distribution at CIPS/DW is rotated clockwise with varying incident light polarization angle $\varphi$. Similar tuning behavior is also observed on the thinner CIPS/PZT heterostructures (Supplemental Fig. S6-S7). These tunable SHG behaviors are more clearly observed from the angle $\theta$ dependent SHG intensity taken on CIPS/DW at different incident light polarization directions [Fig. 4(e)-4(h)]. For the circular patterns shown in Fig. 4(a)-4(d), we define the direction of minimum SHG pointing to maximum SHG (red dashed arrows) as the polarization direction of spatial SHG light generated at CIPS/DW heterointerface, and the angle $\Phi$ between spatial SHG light polarization and *x*-axis as the phase angle of spatial SHG light. Figure 4(i) shows the relationship between spatial SHG light phase angle $\Phi$ and incident light polarization angle $\varphi$. It is interesting to note that the SHG light phase angle almost double that of the incident light polarization angle: $\Phi \approx 2\varphi$, implying that the CIPS/PZT heterostructure can act as an unconventional light polarizer, such as ultrathin half-wave plate. Furthermore, we analyze the SHG light polarization at CIPS/DW heterointerfaces with fixed $\theta$. Figure 4(j) and 4(k) show the SHG intensity versus incident light polarization angle $\varphi$ at the CIPS/DW ($\theta \approx 0^o$) and CIPS/DW ($\theta \approx 180^o$), respectively, all of which exhibit $C_{2v}$-symmetry pattern. For the 9-50 nm CIPS/DW [$\theta \approx 0^o$, Fig 4(j)], SHG reaches the maximum intensity at $\varphi \approx 0^o$ and $180^o$, and minima at $\varphi \approx 90^o$ and $270^o$. The signal at 9-50 nm CIPS/DW [$\theta \approx 180^o$, Fig. 4(k)] exhibits similar angular $\varphi$ dependence, with the pattern rotated by approximately $90^o$. This tuning pattern is expected to be applied to other DWs in both thick and thin CIPS/PZT heterostructures, demonstrating the capability in controlling the SHG light polarization at the nanoscale. Based on the nonlinear electromagnetic theory, the polarization of SHG at the CIPS/DW heterointerface ($\mathbf{P}^{2\omega}_{\mathrm{interface}}$) can be obtained by summing up the second-order polarization vectors for CIPS ($\mathbf{P}^{2\omega}_{\mathrm{CIPS}}$) and PZT DW ($\mathbf{P}^{2\omega}_{\mathrm{DW}}$):[49]

$$\mathbf{P}^{2\omega}_{\mathrm{interface}} = \mathbf{P}^{2\omega}_{\mathrm{CIPS}} + \mathbf{P}^{2\omega}_{\mathrm{DW}} \propto d^{(2)}_{\mathrm{CIPS/DW}} \vec{E}^2(\omega) = (d^{(2)}_{\mathrm{CIPS}} + d^{(2)}_{\mathrm{DW}}) \vec{E}^2(\omega), \tag{4}$$

where $d^{(2)}_{\mathrm{CIPS}}$, $d^{(2)}_{\mathrm{DW}}$, and $d^{(2)}_{\mathrm{CIPS/DW}}$ are second-order $d$-tensors for CIPS, PZT DW, and CIPS/DW heterointerface, respectively. The $d$-tensor for DWs($\theta$) can be obtained from that for PZT domain via the rotation matrix transformation.[48] Then, we can obtain the interfacial SHG intensity at CIPS/DW($\theta$):

$$I_{\mathrm{SHG,CIPS/DW}(\theta)} = \left|\mathbf{P}^{2\omega}_{\mathrm{interface}}\right|^2. \tag{5}$$

For example, the $d$-tensors for DW($\theta = 0^o$) and DW($\theta = 180^o$) can be expressed as:[47]

$$d^{(2)}_{\mathrm{DW}(\theta=0^o)} = -d^{(2)}_{\mathrm{DW}(\theta=180^o)} = \begin{bmatrix} 0 & 0 & 0 & 0 & 0 & d'_{15} \\ d'_{15} & d'_{33} & d'_{15} & 0 & 0 & 0 \\ 0 & 0 & 0 & 0 & d'_{15} & 0 \end{bmatrix}. \tag{6}$$

The interfacial tensors at the CIPS/DW($\theta = 0^o$) and CIPS/DW($\theta = 180^o$) can thus be expressed as:

$$d^{(2)}_{\text{CIPS/DW}(\theta=0^{\text{o}})} = \begin{bmatrix} d_{11} & d_{11} & d_{13} & 0 & d_{15} & d'_{15} \\ d'_{15} & d'_{33} & d'_{15} & d_{15} & 0 & d_{11} \\ d_{15} & d_{15} & d_{33} & 0 & d_{13}+d'_{15} & 0 \end{bmatrix},$$

$$d^{(2)}_{\text{CIPS/DW}(\theta=180^{\text{o}})} = \begin{bmatrix} d_{11} & d_{11} & d_{13} & 0 & d_{15} & -d'_{15} \\ -d'_{15} & -d'_{33} & -d'_{15} & d_{15} & 0 & d_{11} \\ d_{15} & d_{15} & d_{33} & 0 & d_{13}-d'_{15} & 0 \end{bmatrix}. \tag{7}$$

To simplify the calculation, we assume that the SHG intensities from CIPS and PZT DW are the same. In this case, $d'_{15} \approx d_{11}$ and $d'_{15} \approx d'_{33}$. By substituting Equations (7) and (4) into Equation (5), we obtain the interfacial SHG intensity at CIPS/DW ($\theta$ = 0$^{\text{o}}$) and CIPS/DW ($\theta$ = 180$^{\text{o}}$):

$$I_{\text{SHG,CIPS/DW}(\theta=0^{\text{o}})} \propto d_{11}(1+\cos(2\varphi))^2,$$

$$I_{\text{SHG,CIPS/DW}(\theta=180^{\text{o}})} \propto d_{11}(1-\cos(2\varphi))^2. \tag{8}$$

The solid lines in Figs. 4(j) and 4(k) show the modeled SHG results using Equation (8), which are in good agreement with the experimental data.

In summary, we achieve nanoscale modulation of the magnitude and polarization of SHG optical response in layered CIPS, ranging from few-layer to thick-layer dimensions, by interfacing with prepatterned PZT thin films. Combined experimental and theoretical analyses reveal that giant, deterministic modulation of SHG light intensity and polarization is solely determined by the polar alignment between CIPS and ferroelectric DW in PZT. Our study thus provides a new material platform for the rational design of tunable nonlinear optical functionalities, in which the nonlinear optical response could be modulated at the nanoscale through electrical operation or other external control.

**Acknowledgments**

The work at Dalian University of Technology was supported by the National Natural Science Foundation of China (Grant No. 12274051), the Natural Science Foundation of Liaoning Province (Grant No. 2024-MSBA-06), the Liaoning Province Xingliao Talents Plan Project (Grant No.

XLYC2403069), the High-end Foreign Experts Recruitment Plan of China (Grant No. 110000204420258011), the Fundamental Research Funds for the Central Universities (Grant No. DUT24RC(3)060), the Project of the China Scholarship Council to Promote International Cooperation and Training with Russia, Ukraine and Belarus (Grant No. EWXM2509110228), and the National Key Research and Development Program of China (Grant No. 2024YFE0213500). The work at University of Nebraska-Lincoln was supported by the U.S. National Science Foundation (NSF) through Grant No. DMR-2118828 and EPSCoR EQUATE Award No. OIA-2044049, and the UNL Grand Challenges catalyst award.

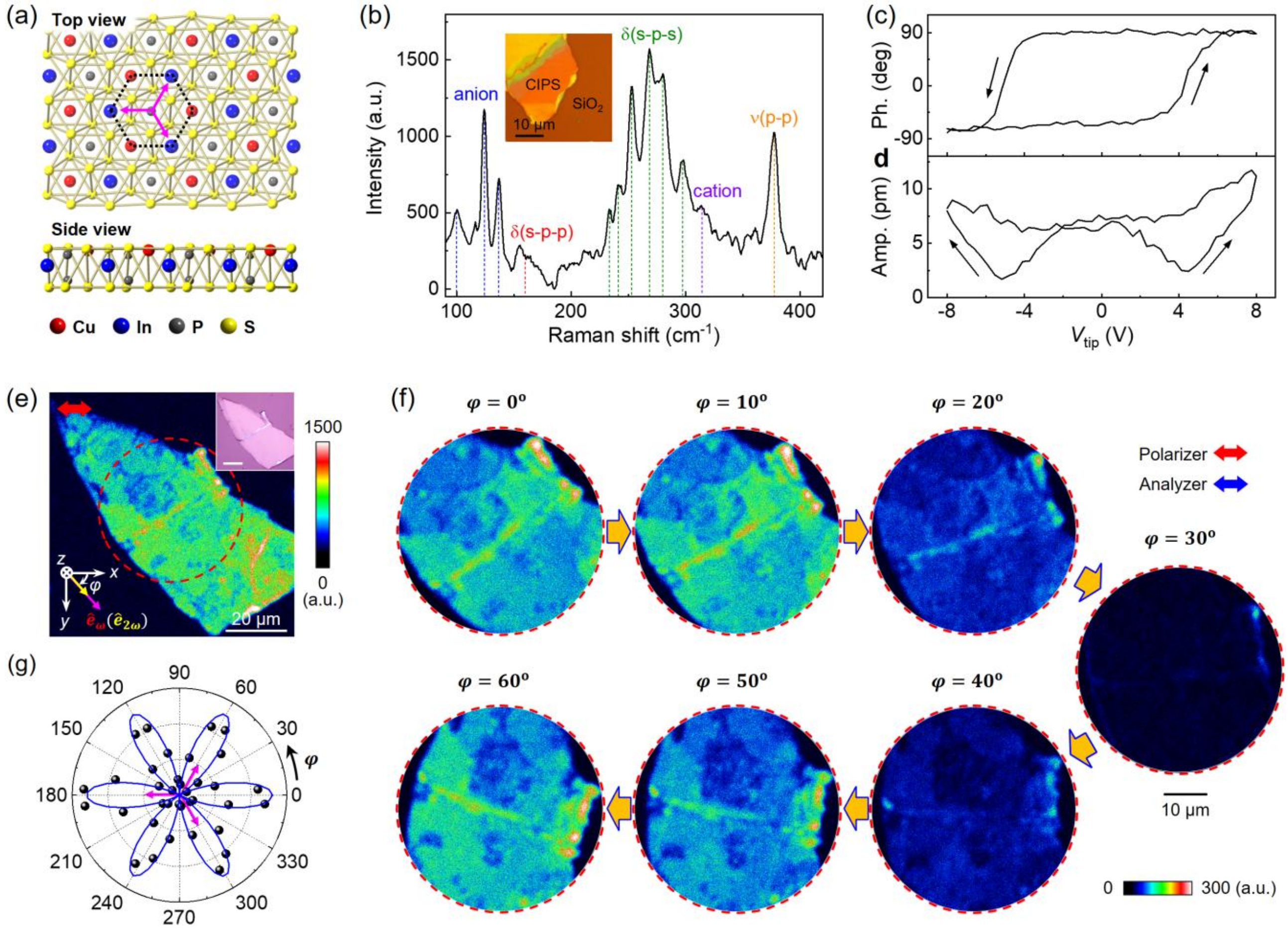

FIG. 1. (a) Top and side schematic views of CIPS crystal structure, where Cu and In atoms form hexagonal in-plane polar lattices. (b) Raman spectrum for an exfoliated CIPS flake (~106 nm) on $SiO_2$/Si substrate (inset). (c, d) PFM switching hysteresis of (c) phase and (d) amplitude responses taken on an exfoliated CIPS flake (~16 nm) on gold substrate. (e-g) SHG taken on an exfoliated CIPS flake (~50 nm). (e) SHG mapping with no analyzer applied. Top inset: optical image. (f) Parallel polarized SHG mapping taken at different sample rotation angles $\varphi$. The red (blue) solid arrows in (e) and (f) mark the incident light (analyzer) polarization. The laboratory coordinate system is shown in (e) bottom inset. (g) Polar plot of parallel polarized SHG intensity as a function of $\varphi$.

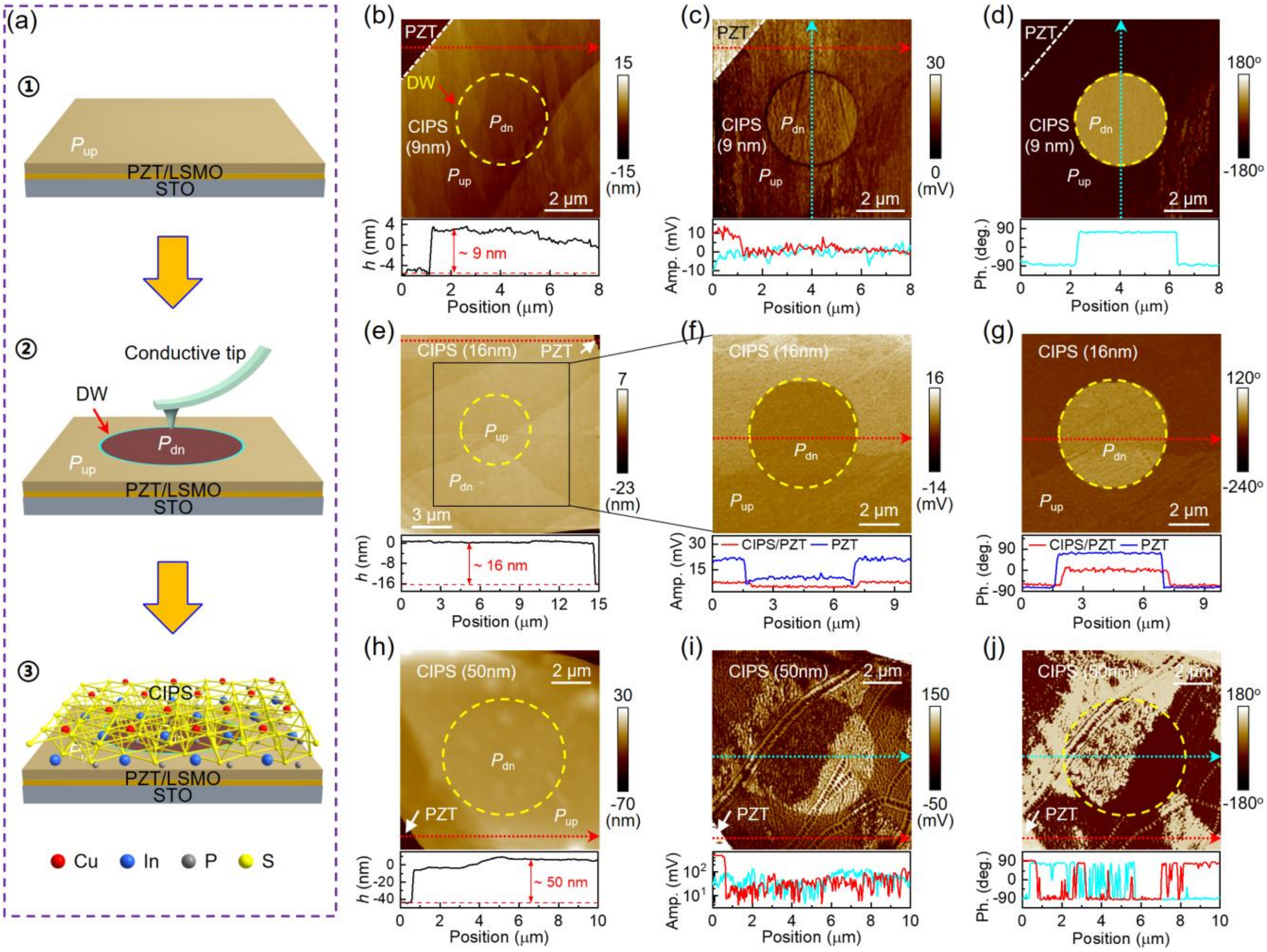

FIG. 2. (a) Schematic workflow for preparing a CIPS flake on PZT with prepatterned circular domain. (b-d) AFM topography (b), PFM amplitude (c), and PFM phase (d) images of a ~9 nm CIPS flake on PZT. (e-g) AFM topography (e), PFM amplitude (f), and PFM phase (g) images of a ~16 nm CIPS flake on PZT. (h-j) AFM topography (h), PFM amplitude (i), and PFM phase (j) images of a ~50 nm CIPS flake on PZT. The yellow dashed lines highlight the DW position. Bottom panels in (b-j): signal profiles along the red and blue dotted lines in the corresponding image.

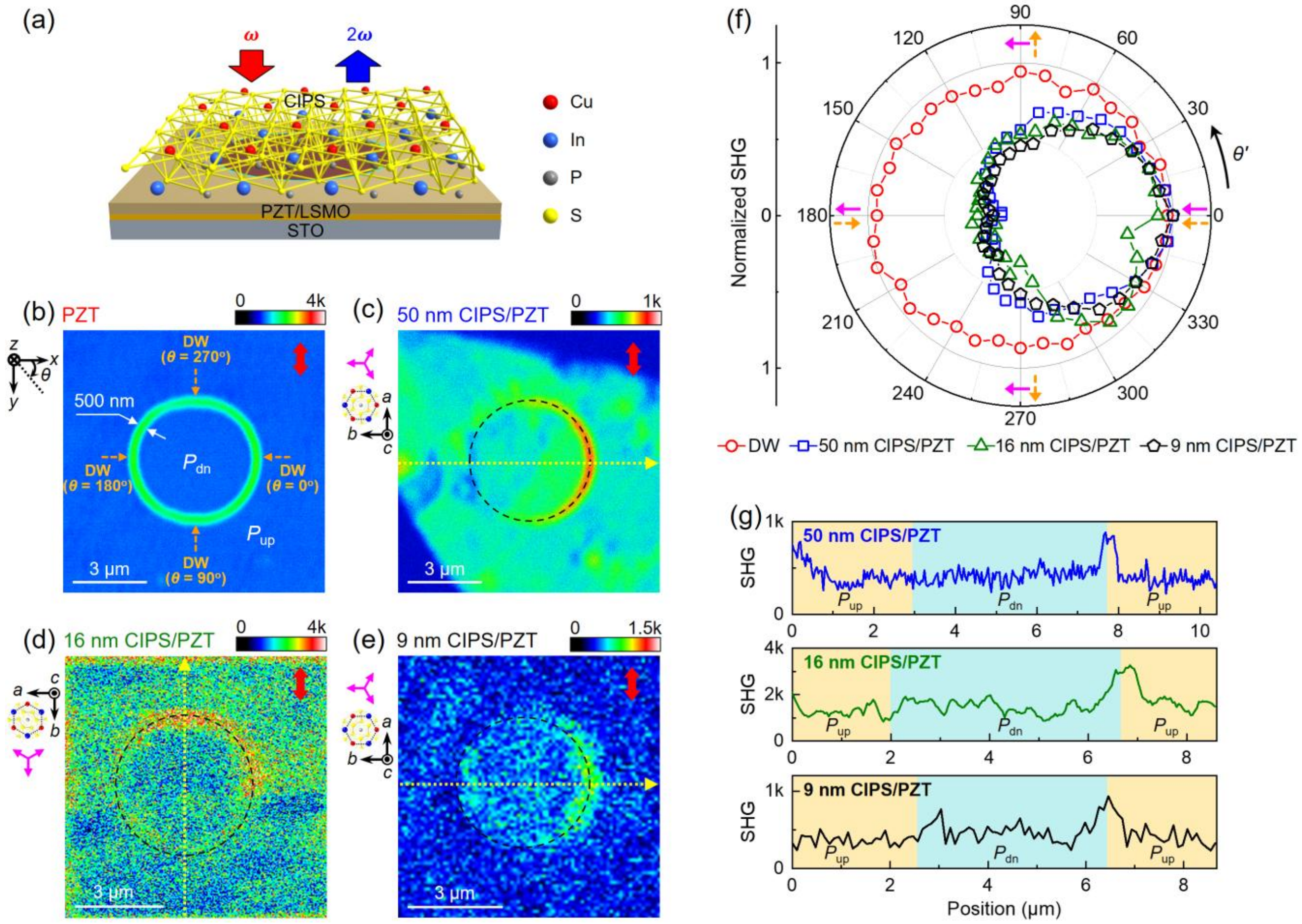

FIG. 3. (a) Schematic of SHG measurement setup. (b-e) SHG mapping of (b) bare PZT, (c) 50 nm CIPS/PZT, (d) 16 nm CIPS/PZT, and (e) 9 nm CIPS/PZT. The laboratory coordinate system and the crystalline orientation of CIPS are shown as insets. The red solid arrows mark the incident light polarization. The yellow dashed (pink solid) arrows mark the polarization directions of DWs (CIPS). Angle $\theta$ is defined as DW position with respect to *x*-direction. The excitation laser power: 20 mW for (b, c) and 25 mW for (d, e). (f) Polar plots of normalized SHG intensity versus $\theta$' taken at DWs on bare PZT in (b) and CIPS/PZT samples in (c-e). Angle $\theta$' is defined as the orientation of SHG signal at DW, and $\theta$' = 0$^{o}$ when SHG signal reaches maximum. Inset: the corresponding $\vec{P}_{\mathrm{DW}}$ (yellow arrows) and $\vec{P}_{\mathrm{CIPS}}$ (pink arrows) directions. (g) SHG intensity profiles along *b*-axes, the yellow dashed lines in (c-e).

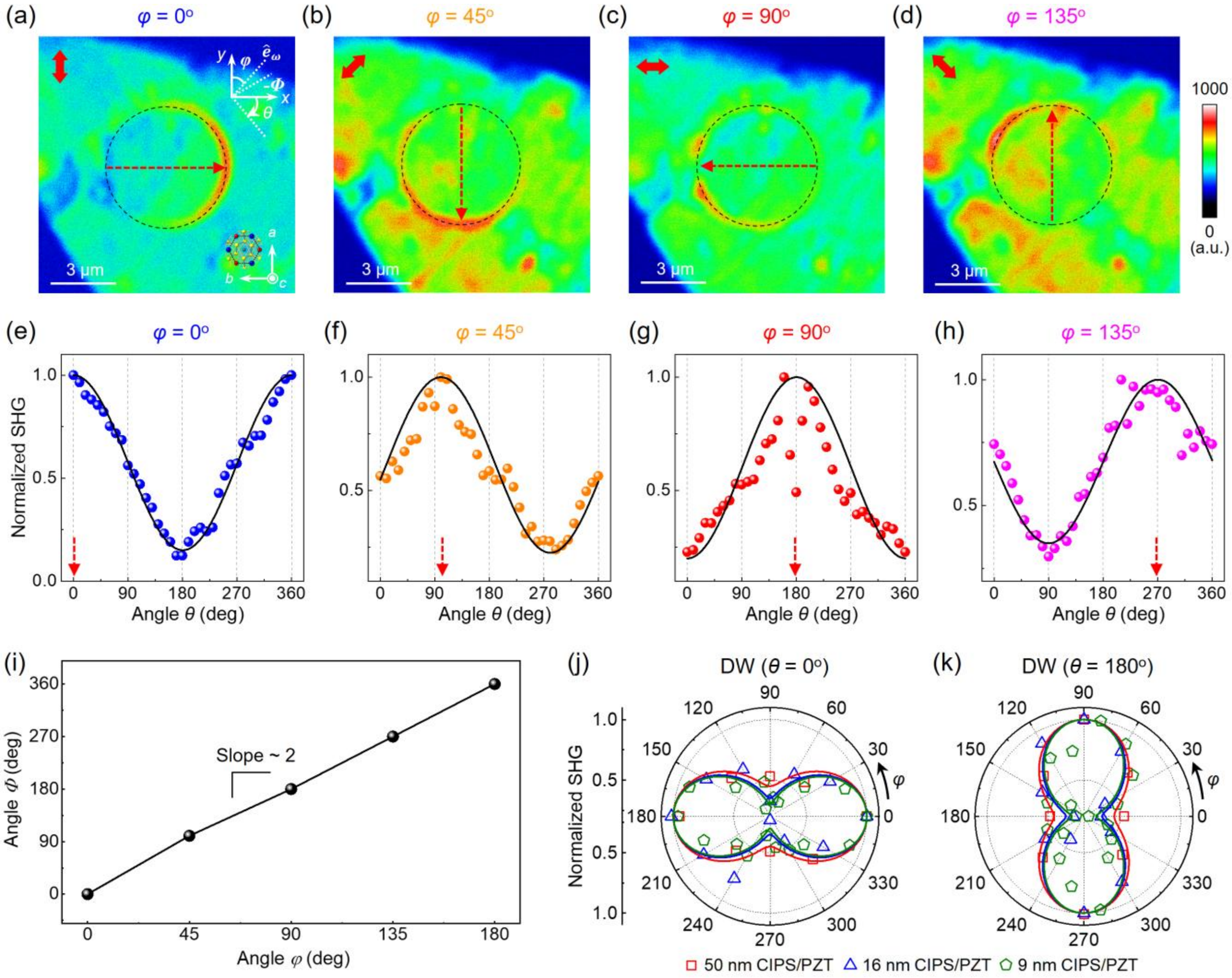

FIG. 4. (a-d) SHG mapping of the 50 nm CIPS/PZT heterostructure taken at different incident light polarization directions: (a) $\varphi$ = 0°, (b) $\varphi$ = 45°, (c) $\varphi$ = 90°, (d) $\varphi$ = 135°. No analyzer is applied. The crystalline orientation of CIPS and the laboratory coordinate system are shown as insets, where $\varphi$ represents the incident light polarization angle, and $\Phi$ represents the spatial SHG light (red dashed arrow) phase angle. The excitation laser power is 20 mW. (e-h) Normalized SHG intensity taken on CIPS/DW versus $\theta$ extracted from (a-d) with fits. (i) Spatial SHG light phase angle $\Phi$ versus incident light polarization angle $\varphi$. (j-k) Polar plots of normalized SHG intensity versus $\varphi$ taken at (j) DW($\theta$ = 0°) and (k) DW($\theta$ = 180°) on 50 nm CIPS/PZT (blue dots), 16 nm CIPS/PZT (green dots), and 9 nm CIPS/PZT (black dots) samples, with fits.